\documentclass[prl,twocolumn,showpacs,preprintnumbers,amsmath,amssymb,floatfix,reprint]{revtex4}
\usepackage{hyperref}
\usepackage{epsfig}
\begin{document}

\title{Parameter-free dissipation in simulated sliding friction}

\author{A. Benassi$^2$, A. Vanossi$^{1,2}$, G.E. Santoro$^{1,2,3}$ and E. Tosatti$^{1,2,3}$}
\affiliation{
$^1$ International School for Advanced Studies (SISSA), Via Beirut 2-4, I-34014 Trieste, Italy \\
$^2$ CNR-IOM Democritos National Simulation Center, Via Beirut 2-4, I-34014 Trieste, Italy \\
$^3$ International Centre for Theoretical Physics (ICTP), P.O.Box 586, I-34014 Trieste, Italy
}

\date{\today}

\begin{abstract}
Non-equilibrium molecular dynamics simulations, of crucial importance in sliding friction, are 
hampered by arbitrariness and uncertainties in the way 
Joule heat is removed. We implement in a realistic frictional simulation a
parameter-free, non-markovian, stochastic dynamics, which, 
as expected from theory, absorbs Joule heat 
precisely as a semi-infinite harmonic substrate would.
Simulating stick-slip friction of a 
slider over a 2D Lennard-Jones solid, we 
compare our virtually exact frictional results with 
approximate ones from commonly adopted empirical dissipation schemes. 
While the latter are generally in serious error, we show that the exact results 
can be closely reproduced by a viscous Langevin dissipation at the boundary layer,
once the back-reflected frictional energy is variationally optimized. 
\end{abstract}


\maketitle
The role of
Molecular Dynamics (MD) simulation in the theory of sliding friction and nanofriction 
can hardly be overestimated~\cite{persson_book,robbin_muser}.
In a typical simulation, a slider is driven by an external force 
over a simulated solid substrate whose atoms, interacting through
realistically chosen interatomic forces, 
vibrate and move according to Newton's law.
Naturally, in order to attain a frictional steady state  
the Joule heat must be removed. Unfortunately, 
a realistic energy dissipation
is generally impossible to simulate reliably,
owing to size (and time) limitations. In a small size
simulation cell, the phonons generated at the sliding interface are 
back-reflected by the cell boundaries, rather than propagated away to properly
disperse the Joule heat. The ensuing problem is a spuriously accumulating 
phonon population in the slider-substrate interface region. The empirical
introduction in the equations of motion of ad-hoc Langevin viscous damping 
terms $-m \gamma_i \dot{q}_i$ (with $m$ and $\dot{q}_i$ the mass and the 
velocity of the $i$-th particle) and of an associated random noise, 
corresponding to some ``thermostat'' temperature $T$~\cite{zwanzig}, 
represents the handiest and commonest solution. However, both this procedure
and the choice of thermostat and damping parameters $\gamma_i$ are vastly arbitrary.
The problem is not just one of principle.
Dry friction generally 
involves stick-slip~\cite{persson_book}, and each slip generates a burst 
of phonons of very specific nature and composition. 
In realistic simulations the overall effect on frictional dynamics of 
the partial back-reflection of this burst must be minimized. 
Unless dealt with, back-reflection will cause the simulated steady state and friction coefficient 
to depend upon unphysical damping parameters. In the Prandtl-Tomlinson model~\cite{vanossi} for instance,
damping is known to modify the kinetics and the friction,
both in the stick-slip regime (including the multiple-slips seen in Atomic 
Force Microscopy~\cite{carpick}) and in the smooth sliding state.
To a lesser or larger extent this lamentable state of affairs is 
common to all MD frictional simulations.

Pursuing a viable solution, one wishes to modify the equations of motion inside a relatively small 
simulation cell, so that they will reproduce the frictional dynamics 
of a much larger 
system, once the remaining variables are integrated out.
Integrating out degrees of freedom is a classic problem, 
largely analysed in literature~\cite{rubin1,zwanzig,we2,kantorovich1,kantorovich2}.
In the context of MD simulation, Green's function methods were formulated for
quasistatic mechanical contacts~\cite{muser}; approaches based on a discrete-continuum 
matching have also been discussed~\cite{luan2,we}.
\begin{figure}
\centering
\includegraphics[width=8.5cm,angle=0]{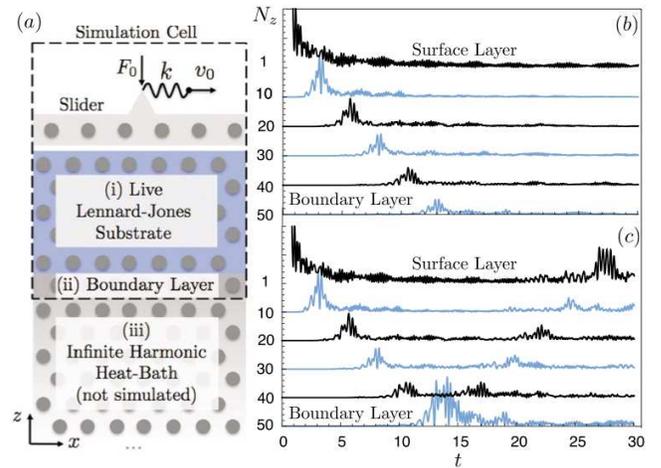}
\caption{(Color online) (a) Sketch of the simulated system. (b),(c) Layer-averaged kinetic 
energy time evolution of a surface-injected phonon burst. Phonon back-reflection is strong without dissipation at the boundary layer (c),  
but accurately canceled once the correct dissipative kernels are included (b).}
\label{figura1}
\end{figure}
%
Among others, a formally exact dissipation
formulation was given early on by Rubin~\cite{rubin1}.
Integrating out $N-1$ atoms in a linear harmonic chain
yields a non-markovian Langevin-like equation of motion of a single 
atom of interest. 
Extensions of this method were
applied to a variety of problems, including the relaxation of impurity molecules in solids~\cite{nitzan1,nitzan2},
and atom-surface scattering~\cite{adelman1,adelman2}.
Detailed formulations suitable for MD simulations were given by Li {\it et al.}~\cite{we2}
and by Kantorovich~\cite{kantorovich1,kantorovich2,kantorovich3}. The idea to variationally minimize 
back-reflection of individual phonons has also been put forward~\cite{we2}.  
However, this body of theory has not yet found its way into realistic 
sliding friction MD simulations, so that neither the importance of
back-reflection errors in dry friction, nor a realistic way to get rid of them
have really been demonstrated. Here we describe how both goals are achieved, 
in a simulated 2D Lennard-Jones system that exhibits a realistic stick-slip. 
We find that 
exact stick-slip friction is indeed different from that of empirical damping schemes. 
However the back-reflected energy 
can be variationally minimized -- although differently 
from Li {\it et al.} --- the resulting approximate dynamics now reproducing 
quite accurately the exact benchmark stick-slip friction.

We focus on a model sliding system made of a semi-infinite 
two-dimensional crystalline substrate with
$N_z$ monatomic layers each 
of $N_x$ atoms, and of a slider made of a
single chain of $N'_x$ atoms. All atoms interact via a Lennard-Jones (LJ) potential (cutoff for
simplicity to first-neighbors). 
The slider is pressed against the substrate by a normal ``load'' force $F_{0}$, and is
driven along $x$ (parallel to the substrate) through a spring of strength $k$, whose end
is pulled at constant velocity $v_0$.
Following similar earlier formulations~\cite{we2,kantorovich1,kantorovich2}, the ideal 
infinitely thick substrate is divided as in Fig.~\ref{figura1}(a), into 
three regions: (i) a live slab comprising $N_z-1$ atomic layers
whose motion is fully simulated by Newton's equations; (ii) the dissipative boundary
($N_z$-th) layer, whose motion includes the effective
non-markovian Langevin terms; (iii) the remaining semi-infinite solid acting 
as a phonon absorber, or heat bath, whose degrees of freedom are 
integrated out, providing the source of effective damping terms in (ii).
In our practical implementation we substitute the full LJ potential within
region (iii) and between (ii) and (iii) with its harmonic approximation. 
This approximation, necessary in order to get
an analytical form for the effective forces on the boundary atoms, 
is all the more accurate the weaker the intensity of frictional phonons. 
In practice, for crystal substrates 
above quantum freezing and below 
the Debye temperature, this level of accuracy can always be attained by a sufficient 
thickness $N_z-1$ of the live simulation cell (i). The harmonic heat bath (iii)
is decoupled by diagonalizing the dynamical matrix $D_{\mu\nu}^{kl}$  ($k,l$ denoting atoms),
obtaining eigenvalues $\omega^2_k$ and eigenvectors $\boldsymbol{\lambda}_k$.
The 
equations of motion
for the boundary layer atoms are \cite{we2,kantorovich1}
\begin{eqnarray} \label{meco}
m\ddot{q}^{i}_{\mu}(t) &=& -\frac{\partial U_{sub}}{\partial q^{i}_{\mu}} -
  m \sum_{j,\nu} \int_{0}^{t} \! ds \; K^{ij}_{\mu\nu}(t-s) \;\dot{q}^{j}_{\nu}(s) \nonumber \\
   && + R^{i}_{\mu}(t) + \sum_{j,\nu} q^{j}_{\nu}(t) \bigg( K^{ij}_{\mu\nu}(0) - D^{ij}_{\mu\nu} \bigg) \;,
\end{eqnarray}
where $i$ and $j$ denote boundary layer atoms,
$\mu$ and $\nu$ indicate $x/z$ components, $U_{sub}$ is the 
LJ interaction between the $i$-th boundary atom and those
inside the simulation cell (i). The second term is non-markovian and non-conservative, 
introducing an effective damping proportional to the velocity of the $j$-th atom,
through a time convolution with the memory kernel functions $K^{ij}_{\mu\nu}(t)$.
Standard kernels are built from the harmonic eigenvalues and eigenvectors of the heat-bath dynamical matrix
and from the coupling vectors $\boldsymbol{\phi}_{\mu}^{i}$ containing the harmonic coupling constants $D_{\mu\mu}^{ik}$ and
$D_{\mu\nu}^{ik}$ of the $i$-th atom of region (ii) with the $k$-th heat-bath atom
$K^{ij}_{\mu\nu}(t) = \sum_{k}
 \bigg[ \frac{(\boldsymbol{\lambda}_k\cdot \boldsymbol{\phi}_{\mu}^{i})
              (\boldsymbol{\phi}_{\nu}^{j}\cdot \boldsymbol{\lambda}_{k})}{\omega^2_k} \bigg] \; \cos{(\omega_k t)}$.
They 
oscillate and decay with time, with power law tails due to the bath 
acoustical phonon branches.
Periodic boundary conditions along the $x$ direction guarantee translational invariance, so
that $K^{ij}_{\mu\nu}(t)$ is a function of $\vert i-j \vert$ only.
As kernels inherit their symmetry properties from those of
the heat-bath dynamical matrix, one can show that $K^{ij}_{\mu\nu}(t)=-K^{ij}_{\nu\mu}(t)$ and $K^{ij}_{\mu\nu}(t)=K^{ji}_{\nu\mu}(t)$.
As $\vert i-j \vert$ grows, $\vert K^{ij}_{\mu\nu}(t)\vert$ decrease, but again not exponentially, 
so that correlations must be included up to large distance~\cite{we2}.
%
By cutting kernels off at time $\tau_c$ one could limit the time-integrals in Eq.~(\ref{meco}),
which need to be calculated at each time step.
By increasing both $\tau_c$ and the live simulation cell size $N_z$,
one refines the dissipation scheme accuracy 
as much as desired, although at increased computational cost.
The third term in Eq.~(\ref{meco}) is a gaussian stochastic force present 
at non-zero temperature, 
responsible for the energy transfer between the heat-bath (iii)
and the live substrate (i), with $\langle R^{i}_{\mu}(t) \rangle=0$, and
%
$\langle R^{i}_{\mu}(t) R^{j}_{\nu}(t')\rangle = m k_B T K^{ij}_{\mu\nu}(t-t') $
%
where brackets denote 
ensemble average, $k_B$ is Boltzmann's constant and $T$ the temperature.
As is well known~\cite{zwanzig}, this relationship fulfils the fluctuation-dissipation theorem --- 
the scheme represents a well defined thermostat in contact with the simulation cell.
The last term in Eq.~(\ref{meco}) 
is the harmonic coupling between atoms $i$-th and $j$-th
within the boundary layer, where the coupling constant $D_{\mu\nu}^{ij}$ is modified
by $K^{ij}_{\mu\nu}(0)$.

As a first step before simulating friction we implement the set 
of equations (\ref{meco}), along with ordinary Newton's equations governing the live cell
in an MD simulation, to test phonon reflection. Starting with the 
equilibrated system ($N_z=30$ close packed 
layers and $N_x=10$ atoms per layer) at temperature $T$,
a test burst of phonons is introduced at the topmost layer and time-evolved.
Results in Fig.~\ref{figura1}(b) show how a relatively thin
$N_z=30$ layer substrate (i+ii) is able to mimick, within the exact dissipation scheme,
the full ideal semi-infinite system (i+ii+iii).
Layer-resolved kinetic energies inside the simulated substrate
show the group of phonons propagating below the surface. Upon
reaching the boundary layer the phonons are perfectly absorbed 
as if they propagated into the (integrated out) semi-infinite crystal (iii).
For comparison, Fig.~\ref{figura1}(c) shows the same phonons
massively back reflected once the memory kernels are removed from the 
boundary layer.

\begin{figure}
\centering
\includegraphics[width=8.5cm,angle=0]{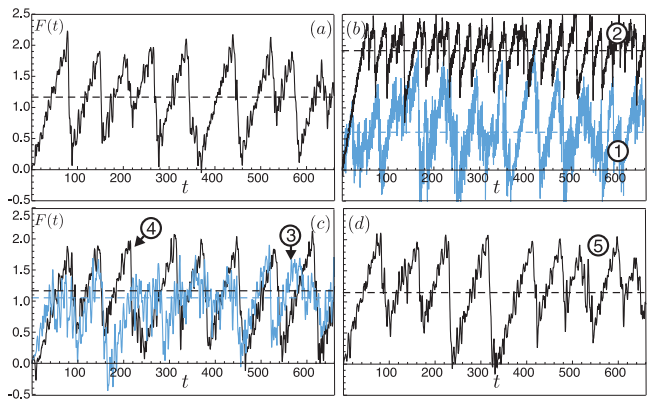}
\caption{(Color online) (a) Simulated friction force $F(t)$
for the ``exact'' non-markovian dissipation scheme of Eq.~(\ref{meco}), 
and [(b),(c) and (d)] for empirical Langevin schemes with a $-\gamma \dot{q}$ damping applied to tip {\bf L1}, whole substrate 
{\bf L2} and bottom layer {\bf L3}. The $\gamma$ value is identified by numbers $1$-$5$ in Fig.\ref{figura3}. Dashed lines: mean value $\langle F \rangle$.}
\label{figura4}
\end{figure}

We next simulate stick-slip sliding friction by driving a slider,
here consisting of a LJ chain of $N'_x=9$ atoms, over the same
live substrate as above
~\footnote{Sliding simulations are performed at temperature $k_BT=0.035$,
roughly
corresponding to $T/T_{melting}=0.06$ (LJ units used throughout).
The load is $F_{0}=10$, the average sliding velocity $v_0=0.01$, and the spring constant $k=5$.
Periodic boundary conditions are applied 
to substrate and 
slider along 
$x$. 
To favor sliding, the strength of the slider-substrate LJ interaction
is reduced from $1$ to
$0.6$. Equations of motion are integrated by a modified velocity-Verlet 
algorithm with a time step of $\Delta t=5\cdot 10^{-3}$. Memory 
kernels are cutoff at $\tau_c =5\cdot 10^{3}$ time-steps,
with good accuracy of the frictional pattern in all its aspects, 
including phonon dissipation, our main concern.}.
The instantaneous friction force is measured by the 
spring elongation $F(t) = k(x_{CM}-v_0 t)$, $x_{CM}$ being the slider center of mass position.
The sawtooth force profile typical of stick-slip friction is obtained (Fig.~\ref{figura4} (a)).
The friction coefficient, obtained by averaging over a stationary sequence of stick-slips 
is $\langle F \rangle/F_{0} = 0.116 \pm 0.002$.
The stick-slip pattern is irregular, with a periodicity similar but not 
exactly matching, a substrate lattice spacing.
The highest spikes signal the forward jump of most slider
atoms, smaller ones involve only about $2/3$  of them.
A measure of the distribution of the spike heights
is the variance of $F(t)$, i.e., $\sigma = \frac{1}{\tau_s \langle F \rangle^2}\int_0^{\tau_s} [F(t)-\langle F \rangle]^2 dt $,
where $\tau_s$ is the total simulation time.
These simulations using the full Eq.~(\ref{meco}), and the corresponding frictional results
can be considered essentially exact, and represent our benchmark reference 
of stick-slip friction with a correct Joule heat removal. The numerical implementation 
of this standard scheme in a generic 3D case will in principle be
possible but time consuming, and far from handy. Long-range 
non-markovian correlations imply a general scaling as $N_{x}^2$, so that
simulations for large-size 3D sliding systems may pose a 
%
%
\begin{figure}
\centering
\includegraphics[width=8.0cm,angle=0]{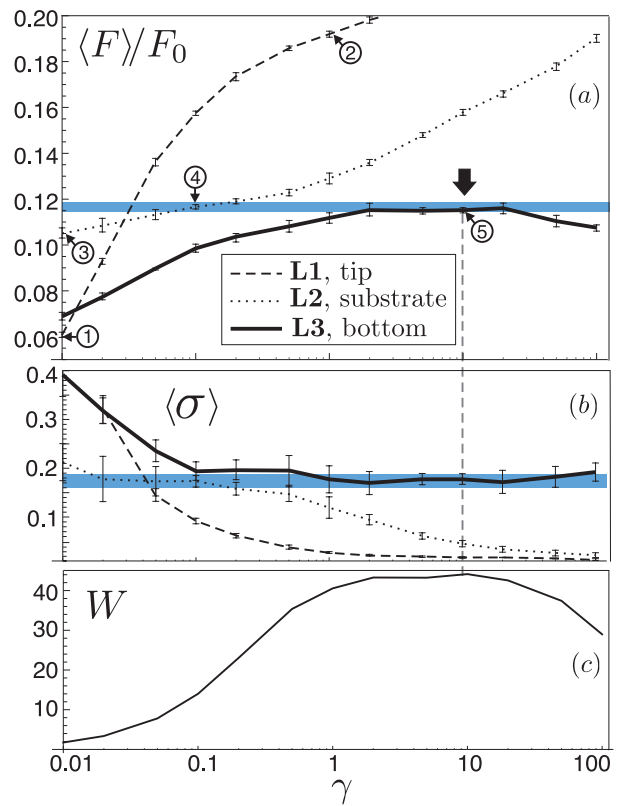}
\caption{(Color online) (a),(b) Friction coefficient $\langle F \rangle/F_0$ and variance $\langle \sigma \rangle$ as a function of
a viscous damping $\gamma$ for different empirical Langevin dissipation schemes, compared with exact non-markovian values (blue stripes). (c) Average internal energy $W$ per substrate atom for the markovian thermostat {\bf L3} divided by the internal energy $W_{exact}$ per substrate atom for the non-markovian scheme (\ref{meco}). The exact and empirical frictional behaviour nearly coincide when the boundary layer 
parameter $\gamma$ is such as to variationally minimize the relative total back-reflected energy $W/W_{exact}$.}
\label{figura3}
\end{figure}
%
practical challenge of parallel computing.

\begin{figure}
\centering
\includegraphics[width=8.5cm,angle=0]{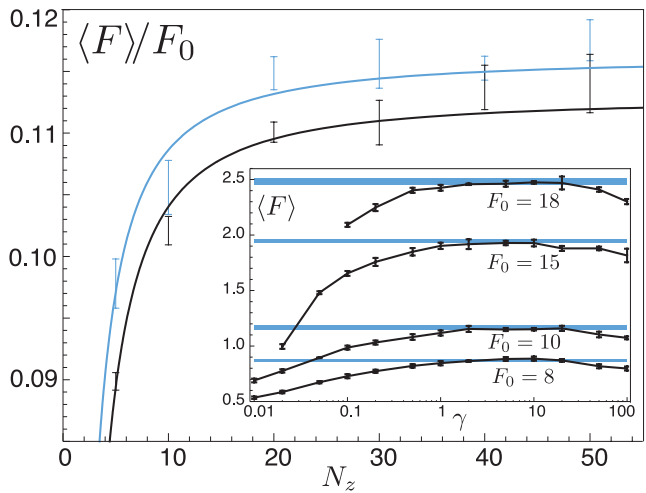}
\caption{(Color online) Friction coefficient as a function of the simulated cell thickness $N_z$ for different driving
velocities $v_0=0.01$ (blue) and $v_0=0.02$ (black) obtained with the non-markovian dissipation scheme.
The inset reports the average friction force for different
loads $F_0$ obtained with the non-markovian approach (blue stripes) and the comparison with the markovian Langevin scheme {\bf L3} at different $\gamma$ (black curves).}
\label{figura5}
\end{figure}

The virtually exact frictional dynamics 
just obtained can now be compared with 
empirical Langevin schemes commonly adopted in friction 
simulations, where a viscous damping $-\gamma \dot{q}$ is arbitrarily applied to the motion of some atoms in the system,
for instance to slider atoms ({\bf L1}, dashed line in Fig.\ref{figura3}) 
or to all substrate atoms ({\bf L2}, dotted line), 
or to the boundary atomic layer only ({\bf L3}, solid line).
As anticipated, 
the stick-slip friction simulated with all 
these empirical 
schemes depends vastly on $\gamma$, and generally 
deviates systematically from the correct benchmark. 
The closest agreement is obtained in case {\bf L3} where 
%
%
%
the viscous damping $-m\gamma \dot{q}^i_{\mu}(t)$ is applied to the $i$-th boundary layer atom motion, 
$m \ddot{q}^{i}_{\mu}(t)=-\frac{\partial U}{\partial q^i_{\mu}} - m \gamma \dot{q}^i_{\mu}(t) + R^i_{\mu}(t)$,
with the appropriate
gaussian stochastic force $R_i(t)$ with $\langle R^i_{\mu}(t) \rangle = 0$ and
$\langle R^i_{\mu}(t)R^{j}_{\nu}(t') \rangle = 2 m k_B T \gamma \delta_{\mu,\nu}\delta_{i,j}\delta(t-t')$.
Since the parameter $\gamma$ is adjustable, we may seek to optimize it in order to approximate 
the exact results.
%
Li {\it et al.} ~\cite{we2} considered  variationally minimizing a group-velocity weighted
phonon reflection. In the actual friction simulation, the steady state internal
energy increase $W = \langle E \rangle - \langle E_0 \rangle $ over the equilibrium
value $E_0$ is easily computable. Our proposed scheme is to variationally minimize $W$,
a quantity generally positive and proportional to the Joule frictional heat.
In absence of back-reflection $W$ is minimal, whereas partial energy 
reflection at the boundary will cause Joule heat to artificially accumulate
in the slab and increase $W$, so that for any arbitrary dissipation scheme $W/W_{exact} > 1$. 
As a function of $\gamma$, a variational minimum of  $W/W_{exact}$ occurs 
because back-reflection of frictional phonons is large {\it both} when
the boundary layer damping $\gamma$ is too small and too large.
This ratio is  shown in Fig.~\ref{figura3}(c). 
The agreement between the exact frictional results, where no phonons are back reflected, and 
the variational one at $\gamma=10$ is excellent. (We note incidentally that 
at the optimal $\gamma$ the friction coefficient also peaks).
To confirm the variational result, we changed system parameters, including 
sliding velocity, and load. 
The variable load results of Fig.~\ref{figura5} (inset) show that the coincidence of optimal and exact friction is systematic.
Fig.~\ref{figura5} also shows a dependence of calculated friction upon the thickness of the simulated substrate portion (i+ii),
converging for sufficiently large $N_z$. The required thickness of a few tens of layers depends on velocity,  
and is reached once the substrate inertia grows large enough to stop interfering spuriously with the frictional dynamics.

In conclusion, we demonstrated frictional MD simulations with a nonmarkovian dissipation,
enacting the correct disposal of generated phonons, even in the relatively violent
stick-slip case. Using the virtually exact frictional results so obtained as a reference, 
we benchmarked common empirical viscous dissipation schemes, finding them generally wanting.
There exists however an optimal viscous damping $\gamma_{opt}$ which, once applied to 
the cell boundary layer, yields results nearly indistinguishable from the exact ones. The optimal damping 
parameter variationally minimizes the total back-reflected energy at the cell boundary, and can be 
identified even without any exact reference calculation. This optimal damping scheme, which unlike
the exact one does not require a knowledge of the substrate vibrational properties, is a good candidate for adoption in 
future practical MD frictional simulations.\\

A discussion with L. Kantorovich is gratefully acknowledged. This work is part of Eurocores Projects 
FANAS/AFRI sponsored by the Italian Research Council (CNR), and of FANAS/ACOF. It is also partly 
sponsored by The Italian Ministry of University and Research, through a PRIN/COFIN contract.

\end{document}